\documentclass[10pt]{spie}  

\usepackage{amsmath,amsfonts,amssymb}
\usepackage{graphicx}
\usepackage{tabularx}
\usepackage{subcaption}
\usepackage[colorlinks=true, allcolors=blue]{hyperref}
\usepackage{aas_macros}

\title{The Atacama Large Aperture Submillimeter Telescope: Key science drivers  }

\author[a]{Joanna Ramasawmy}
\author[a]{Pamela D. Klaassen}
\author[b]{Claudia Cicone}
\author[c]{Tony K. Mroczkowski}
\author[d]{Chian-Chou Chen}
\author[e]{Thomas Cornish}
\author[f]{Elisabete da Cunha}
\author[c]{Evanthia Hatziminaoglou}
\author[g,h]{Doug Johnstone}
\author[i]{Daizhong Liu}
\author[j]{Yvette Perrott}
\author[b]{Alice Schimek}
\author[c]{Thomas Stanke}
\author[b]{Sven Wedemeyer}

\affil[a]{UK Astronomy Technology Centre, Royal Observatory Edinburgh, Blackford Hill, Edinburgh United Kingdom }
\affil[b]{Institute of Theoretical Astrophysics, University of Oslo, PO Box 1029, Blindern 0315, Oslo, Norway}
\affil[c]{European Southern Observatory, Karl-Schwarzschild-Str. 2, 85748 Garching bei München, Germany}
\affil[d]{Academia Sinica Institute of Astronomy and Astrophysics (ASIAA), No. 1, Sec. 4, Roosevelt Road, Taipei 10617, Taiwan}
\affil[e]{Department of Physics, Lancaster University, Bailrigg, Lancaster, LA1 4YB, UK}
\affil[f]{International Centre for Radio Astronomy Research, University of Western Australia, 35 Stirling Hwy, Crawley, WA 6009, Australia}
\affil[g]{ National Research Council of Canada, Herzberg, Astronomy and Astrophysics Research Centre, 5071 West Saanich Road, V9E 2E7 Victoria (BC), Canada}
\affil[h]{Department of Physics and Astronomy, University of Victoria, Victoria, BC V8P 5C2, Canada}
\affil[i]{Max-Planck-Institut für extraterrestrische Physik, Gießenbachstraße 1, 85748, Garching b. München, Germany}
\affil[j]{School of Chemical and Physical Sciences, Victoria University of Wellington, PO Box 600, Wellington 6140, New Zealand}

\authorinfo{Further author information: \\J. Ramasawmy: E-mail: joanna.ramasawmy@stfc.ac.uk, Telephone: (+44) 0131 6688 389}
\newcommand{\kms}{km s$^{-1}$}

\begin{document} 
\maketitle

\begin{abstract}
The Atacama Large Aperture Submillimeter Telescope (AtLAST) is a concept for a 50m class single-dish telescope that will provide high sensitivity, fast mapping of the (sub-)millimeter sky. Expected to be powered by renewable energy sources, and to be constructed in the Atacama desert in the 2030s, AtLAST’s suite of up to six state-of-the-art instruments will take advantage of its large field of view and high throughput to deliver efficient continuum and spectroscopic observations of the faint, large-scale emission that eludes current facilities. 

The AtLAST design study project is currently supported by a Horizon 2020 grant aimed at studying the governance, telescope design, site selection, telescope operations, sustainable energy supply, and science drivers of the future AtLAST observatory. With quantified and specific science goals, we can begin to place technical specifications on the telescope and its instrumentation. As a first step in this process, we conducted a consultation on potential AtLAST science with the global (sub-)millimeter astrophysics community. The consultation involved nearly 100 scientists based in 22 countries, and the resulting 28 use cases indicate the breadth of transformational science that such a high-throughput facility could make possible: from exploring the prebiotic molecular chemistry of comets in our own Solar System, detecting the extended, diffuse cold gas in the circumgalactic medium of both our own and distant galaxies, to detailed measurements of the thermal, kinetic, and relativistic Sunyaev-Zeldovich effect and mapping of large-scale structure. Already these science cases define some core requirements for AtLAST’s instrumentation: wide bandwidths, multichroic observations, high spectral resolution, fast mapping and a large field of view. Further refinement of these is planned over the course of the current EU-funded project, resulting in detailed case studies of the telescope and instrumentation requirements needed by the community to deliver a next-generation submillimeter observing facility.  
\end{abstract}


\section{INTRODUCTION}
\label{sec:intro}  

In this article, we present the preliminary scientific output from the the ongoing AtLAST design study, specifically focusing on the science goals and the corresponding telescope and instrumentation requirements. The AtLAST concept was introduced in a previous paper\cite{2020SPIE11445E..2FK} which detailed the proposal and planning for the project: to design a 50m single-dish next generation submillimeter telescope with high throughput and large field of view (FoV). Now that the project is underway, we can report on the initial results from the science work package. This part of the study is continuously evolving, and the current results encompass to the first year of progress. As the project continues we expect to derive more robust science themes from the community, and as such the science motivation will evolve hand-in-hand with the development of the telescope design and instrumentation goals over the remaining two years of the design study.

In Section~\ref{sec:state-of-the-art}, we describe the state-of-the-art of submillimeter observations, and explain the need for a new facility. In Section~\ref{sec:study}, we briefly outline the broader design study, before delving into the preliminary science driver results in Section~\ref{subsec:science_drivers}. A summary of the requirements placed on the telescope by these science goals, and initial suggestions for generalized instrumentation goals, are outlined in Section~\ref{sec:requirements}. Finally, we detail the path forward for the design study in Section~\ref{sec:future}.

\section{State-of-the-art}
\label{sec:state-of-the-art}
Existing submillimeter and millimeter astronomical observing facilities fall into two categories, offering two distinct methods of observation: single dish telescopes and interferometers.

Current single-dish facilities with moderately large dishes include (but are not limited to) the Atacama Pathfinder Experiment (APEX\cite{2006A&A...454L..13G}, 12m diameter), the James Clerk Maxwell Telescope (JCMT, 15m), and a number of 12m ALMA prototype telescopes such as the Large Latin American Millimeter Array (LLAMA), the Greenland Telescope, and the UArizona ARO 12-meter Telescope. We highlight some notable results from APEX and JCMT as key instruments that can reach the shorter wavelength regime ($<1$mm). 

APEX, as the pathfinder for the Atacama Large Millimeter/Sub-millimeter Array (ALMA), takes advantage of the excellent atmospheric conditions at the Chajnantor plateau enabling observations in the THz frequency regime. A multi-instrument facility telescope, it has also acted as a test bed for new instrumentation concepts such as CONCERTO\cite{2020A&A...642A..60C}, a LEKID detector spectrometer. The Large APEX BOlometer CAmera (LABOCA)\cite{2009A&A...497..945S} has produced the most sensitive submillimeter observations to date at $870\mu$m, including the detection of populations of dusty star-forming galaxies at high redshift\cite{2009ApJ...707.1201W, 2011MNRAS.415.1479W}, as well as some of the most comprehensive surveys of our Galaxy\cite{2009A&A...504..415S,2021MNRAS.500.3064S}.

The JCMT has been the workhorse telescope of extensive sky surveys, with both spectroscopic and photometric instrumentation, and resulting observations have underpinned our understanding of the submillimeter sky: tracing the morphology, chemistry and polarization of star-forming regions in our own galaxy \cite{1999ApJ...510L..49J, 2013ApJ...767..126S, 2019ApJ...880...27P, 2019MNRAS.485.2895E}, mapping the multi-phase gas and dust in galaxies both nearby \cite{2009ApJ...693.1736W, 2019MNRAS.486.4166S} and at high redshift \cite{2017MNRAS.465.1789G}, and probing the chemistry of planetary atmospheres in our own solar system \cite{2021NatAs...5..655G}.
Single-dish telescopes like the JCMT are also uniquely poised to undertake time-variable observations, detecting transient phenomena such as protostellar accretion and stellar flares \cite{ 2021ApJ...920..119L, 2021ApJ...920..119L}.

At larger apertures ($>25$m), there are few facilities that can reach submillimeter wavelengths: these include the IRAM 30m telescope\cite{1987A&A...175..319B} and the Large Millimeter Telescope (LMT, 50m). Figure~\ref{fig:telescopes} shows a comparison of telescope dish diameter and FoV of several existing single dish facilities with AtLAST. Perhaps most importantly it is the \textit{combination} of these two parameters that makes AtLAST stand apart, with the resulting throughput shown in a log$_{10}$ scale in the right hand panel of Figure~\ref{fig:telescopes}, which determines the number of beams on sky.\footnote{See the memo on the optical configuration of AtLAST: \href{https://www.atlast.uio.no/documents/memo-series/memo-public/basic-layout-options-v2.pdf}{Basic Layout Options} by R.~Hills}

While single dish telescopes can map the (sub-)mm sky, they are limited in both resolution and sensitivity by dish size. With large beam sizes, confusion noise dominates and becomes a limiting factor on sensitivity, so higher resolution (i.e. a larger aperture) lowers the confusion limit and thus allows for deeper observations.
The largest aperture facilities capable of reaching the highest resolutions, and therefore sensitivities, feasible with single dishes are limited in wavelength range. There are only two telescopes with dishes larger than 25m -- the LMT and the IRAM 30m -- that have been able to observe $>116$ GHz, and even in optimal conditions can rarely observe up to 350 GHz, which in turn limits their spatial resolution. 
This restricts observations to bands that exclude high frequencies necessary for many fields of study, such as detailed component separation of the thermal, relativistic, and kinetic Sunyaev-Zeldovich (SZ) effects \cite{1972CoASP...4..173S,1980MNRAS.190..413S,2019SSRv..215...17M} from the lensed and cluster-centric dusty source populations, the detection of high-J transition CO, [CI] and fine structure lines across a wide range of redshifts, and constraining the dust spectral energy distribution (SED) of high-redshift galaxies. AtLAST will be the \textit{only} telescope with large aperture that is proposed to reach frequencies $< 400$GHz, enabling these kinds of studies.

Several smaller diameter, wide-FoV single dish facilities such as the Atacama Cosmology Telescope (ACT), CMB-S4, Fred Young Submillimeter Telescope (FYST\cite{2018SPIE10700E..5XP}, formerly CCAT-prime), Simons Observatory, POLARBEAR and the South Pole Telescope (SPT) have been constructed or are in development with the express purpose of mapping the CMB on large scales. These telescopes have relatively low resolution, but extremely large FoV (up to 8 degrees$^2$) allowing them to measure fluctuations in temperature and polarization over large areas of sky; however, they are of limited use for many other science goals as their large beam sizes (e.g. $>1'$) lead to a high confusion limit. It is also worth noting that, despite their large FOVs and survey-specific design, AtLAST will feature a throughput ($A \Omega$)  $\gtrsim 5\times$ larger than any of the aforementioned small aperture telescopes (see Figure~\ref{fig:telescopes}).

(Sub-)Millimeter interferometers such as NOEMA, the SMA and ALMA\cite{2009IEEEP..97.1463W} have allowed us to resolve sub-arcsecond angular scales and reveal the cold Universe in detail and investigate the chemistry and structure of objects in the Universe individually or in targeted surveys.
Over the last decade, ALMA observations have led to numerous firsts and breakthroughs: imaging the protoplanetary disks of young solar analogs \cite{2015ApJ...808L...3A, 2016ApJ...828...46A}, detecting oxygen and other emission lines in a galaxy at redshift $z > 7$ \cite{2022ApJ...929..161W}, and mapping the molecular gas of nearby star-forming and typical main-sequence galaxies in the young universe \cite{2021ApJS..257...43L}, as well as playing a key role in the Event Horizon Telescope observations of the supermassive black holes at the heart of M87\cite{2019ApJ...875L...1E} and our own galaxy\cite{2022ApJ...930L..12A}.
By design, current (sub-)millimeter interferometers like ALMA have a small FoV, which makes mapping any significantly sized area prohibitively time-consuming.  They are also only sensitive to emission on scales inversely proportional to their baseline separations; making them insensitive to large-scale emission -- the so-called ``missing flux'' problem. Thus, interferometers tend to focus on the follow-up of known targets for detailed study, which relies on the quality of large-area, lower resolution mapping available at (sub)mm wavelengths.  This leads to a key problem of ``source starvation'': the exquisite resolution afforded by interferometric facilities allows us to study those bright, rare sources in detail, but without sensitive large-area submillimeter maps we will run out of known sources to study.  

Current (sub-)mm single dish facilities have relatively high confusion limits which hampers how well populations (whether they be proto-stellar or galaxy) can be characterized. Comprehensive mapping at higher sensitivities and with greater resolution will ensure that interferometers can be used to delve into the fainter objects in these populations. For example, in the distant universe, current confusion limits constrain single-dish surveys to quantifying galaxies with luminosities $>10^{12}$L$_\odot$; we can only get a census of the most extreme galaxies. With AtLAST, we push the confusion limits down below the cutoffs expected for the general galaxy population at high-redshift. With the proposed sensitivity and resolution of AtLAST, we will be able to make the first complete, cosmic-variance free, volume-limited census of the distant Universe.

The missing-flux problem means that imaging extended, diffuse submillimeter emission, if too faint to be detected by existing single dish telescopes, is currently impossible. The ability to detect this emission would allow us to probe the chemistry of low surface brightness Universe: for example, investigating the role in galaxy evolution of the gas and dust in the circumgalactic medium\cite{2019BAAS...51c..82C}.

It is clear that there exists a gap in parameter space covered by existing submillimeter observatories: the ability to observe at sufficiently high spatial resolution to reach sensitivities that allow us to map faint sources, and detect the diffuse emission invisible to current telescopes; reaching high frequencies and with the combination of both large aperture and FoV required for high throughput enabling efficient mapping of large sky areas. Several key science questions, that we begin to outline in this document, rely on such observations and are impossible to address without a dedicated multi-instrument, high throughput facility such as AtLAST.

\section{The AtLAST design study}
\label{sec:study}

The AtLAST design study, a 3.5 year Horizon 2020 program, aims to create a fully developed concept for the next-generation submillimeter single dish observing facility. Over the course of the study, work packages focus on governance, telescope design, site selection, telescope operations, sustainable energy supply and science drivers to propose a design for AtLAST. Progress is underway in each of these work packages, and updates and memos can be found on the AtLAST website\footnote{\href{https://www.atlast.uio.no/}{www.atlast.uio.no}}. While here we are focusing on the science drivers for the telescope, we provide a brief summary of the design study below.
 
\begin{figure}
        \centering
        \includegraphics[width=\textwidth]{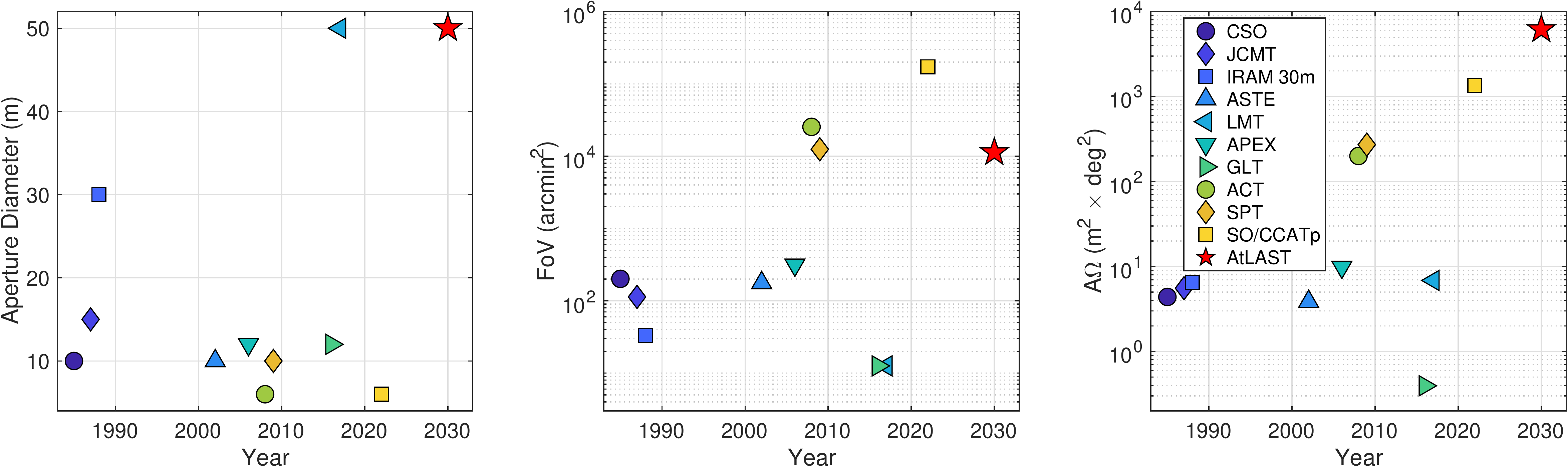}
        \caption{From left to right, plots show the aperture diameter, field of view, and resulting throughput A$\Omega$ of the planned design for AtLAST (denoted with a red star) relative to other single-dish observatories. While there are other telescopes of comparable aperture or FoV, it is the combination of the two that give AtLAST its unique position in the thirdmost plot. Ultra wide-field telescopes such as CCATp begin to reach the throughput, but at much lower resolution and therefore sensitivity, with AtLAST reaching around 5000 times the throughput of comparably sized dishes like the LMT.
        This very high throughput at moderate resolution and deep sensitivity is key to answering the science questions outlined below.
        }
        \label{fig:telescopes}
    \end{figure}

    The main telescope characteristics can be found in Table 1 of our previous paper\cite{2020SPIE11445E..2FK}, and we summarize them briefly here. Key parameters include a 50m aperture and a 2 deg FoV. This combination of large aperture and FoV sets AtLAST well ahead of other existing and upcoming facilities in terms of throughput, making it the most efficient tool for mapping the submillimeter sky at moderate angular resolution ($\sim 1.8^{\prime\prime}$ at 450$\mu$m / $\sim 5^{\prime\prime}$ at 1.1mm). Figure~\ref{fig:telescopes} shows this in the context of other existing and upcoming single-dish submillimeter telescopes: AtLAST's throughput exceeds that of telescopes with comparable dish size by over two orders of magnitude.
    
    The planned site for AtLAST is the Chajnantor Plateau, at an altitude of around 5100 m above sea level. This site has excellent atmospheric transmission (see Figure 3 from Ref.~\citenum{2020SPIE11445E..2FK}), allowing for observations at high frequency ($> 500$ GHz). The exact location is still being determined by the site selection work package, with considerations for the weather conditions, access routes, and development of infrastructure at potential locations.
    
   A work package dedicated to developing more sustainable telescope energy supply will assess the possibilities of providing power via combinations of a photovoltaic park, batteries and novel hydrogen storage methods, combined with other resources like diesel generators.
    
    While at this stage in the design study the scope does not include the development of specialized instrumentation, we are undertaking reviews of the state-of-the-art of submillimeter detectors and instrumentation as a guide for the potential instrumentation goals, together with both the telescope design and science drivers work packages.

\section{Key science drivers}
    \label{subsec:science_drivers}
        
\begin{figure}
        \centering
        \includegraphics[width=\textwidth]{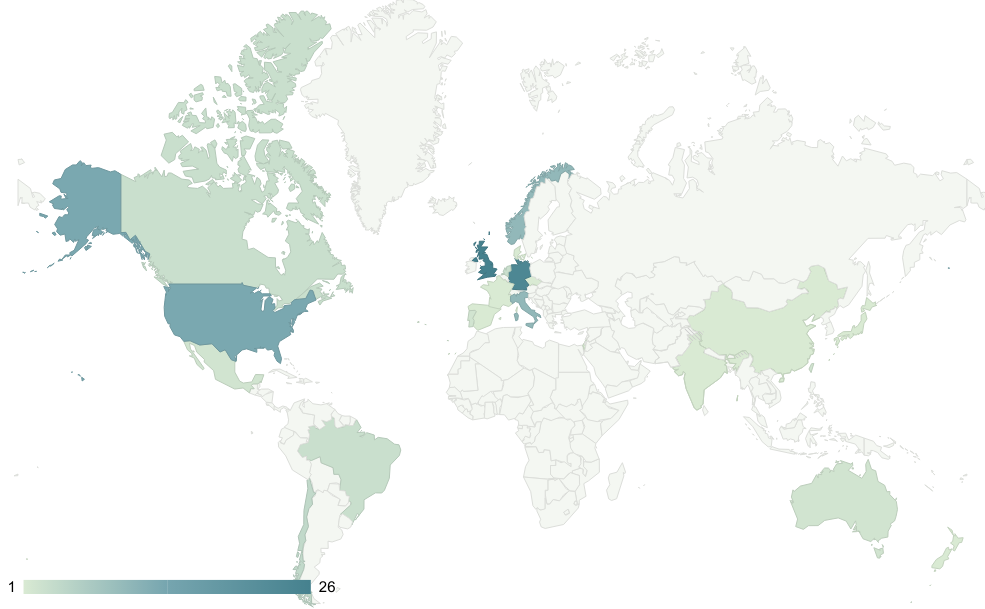}
        \caption{Map showing the distribution of contributors to the development of the AtLAST science cases.}
        \label{fig:world_map}
    \end{figure}

    One year into the Horizon 2020 funded study, progress has been made on capturing the research needs of the broader astronomy community and developing science goals that cover the full breadth of cutting-edge astronomy research, from cometary chemistry to the largest scale structures in the Universe.
    The science goals working package has now engaged with over 100 scientists from $\sim22$ countries (see Figure~\ref{fig:world_map}).
    
    Following a call for initial science use cases in mid-2021, we received 28 submissions\footnote{See summary of submitted use cases here: \href{https://www.atlast.uio.no/design-study/wp6-science/atlast_usecase_summary.pdf}{AtLAST Science Use Cases}} covering a range of astrophysical topics, which we have divided into four science categories: The Sun \& Solar System, The Milky Way, Nearby Galaxies, and Distant Universe \& Cosmology.
    These use cases serve as a starting point in the development of a scientifically justified set of requirements for AtLAST at the end of the design study period.
    They will also eventually guide the choices and development of instrumentation for AtLAST.
    
    Each theme contains several key science goals which require both large-scale surveys and smaller more targeted investigations. This is in line with the discussion in  Reference~\citenum{2020SPIE11445E..2FK} that it is crucial that any upcoming large submillimeter facility is not simply a survey instrument but allows for PI-driven science that can access the unique parameter space such a telescope provides, and we illustrate that there are a wealth of research questions presenting the need for such observing strategies.
    
    \begin{figure}[h]
         \centering
         \begin{subfigure}[b]{0.45\textwidth}
             \centering
             \includegraphics[width=\textwidth]{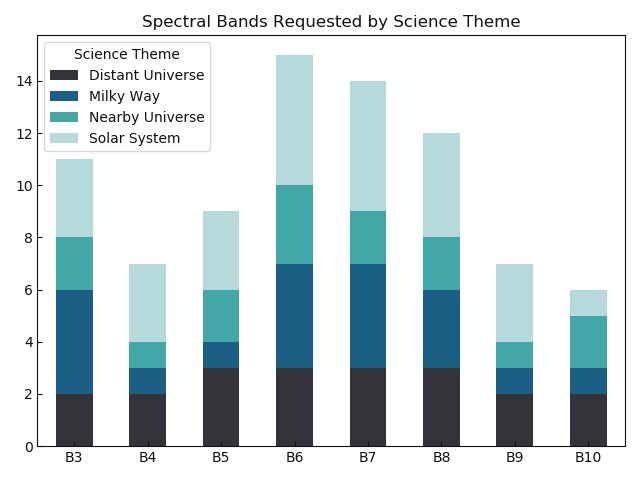}
             \caption{Spectroscopic Bands requested by Science Theme}
             \label{fig:spec_by_sci_theme}
         \end{subfigure}
         \hfill
         \begin{subfigure}[b]{0.45\textwidth}
             \centering
             \includegraphics[width=\textwidth]{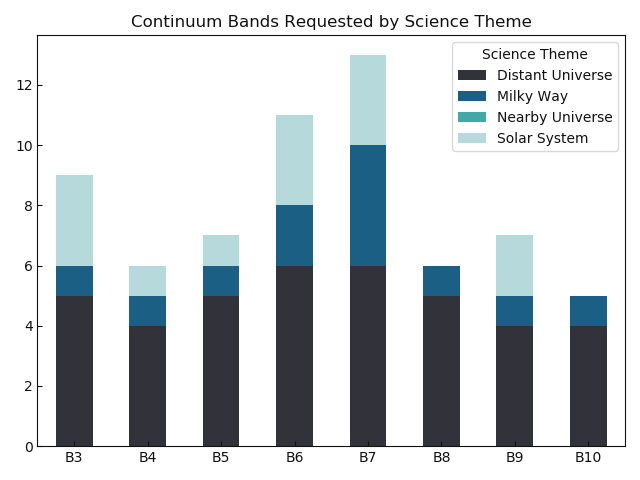}
             \caption{Continuum Bands requested by Science Theme}
             \label{fig:cont_by_sci_theme}
         \end{subfigure}
            \caption{Spectroscopic and Continuum bands requested by Science Theme. The labeling on the x-axis corresponds to the canonical ALMA bands defined in each spectroscopic window which can be found in the ALMA technical handbook\cite{cortes_paulo_2021_4612218} and Table~\ref{tab:alma}.}
            \label{fig:Req_by_theme}
    \end{figure}

    From these initial use cases we extracted 81 requests for spectroscopic observations and 64 requests for continuum. 
    While the telescope requirements most stretched by each science use case are discussed in greater detail in the next sections, we show in Figure \ref{fig:Req_by_theme} the pressures on the different (sub-)mm windows by science theme. Here we refer to the atmospheric windows for observation by their corresponding ALMA frequency bands\cite{cortes_paulo_2021_4612218} due to their familiarity and equivalence to windows accessible to AtLAST, a description of which can be found in Table~\ref{tab:alma} below.
    
    \begin{table}[h!]
    \centering
    \caption{Mapping of ALMA Observing bands to their Atmospheric windows}
    \begin{tabular}{l|cc}
    \hline \hline
         ALMA band  & Wavelength & Frequency  \\
                    & (mm)       & (GHz) \\
        \hline
         3          & 2.6-3.6   & 84-116\\
         4          & 1.8-2.4   & 125-163 \\
         5          & 1.4-1.8   & 163-211 \\
         6          & 1.1-1.4   & 211-275 \\
         7          & 0.8-1.1   & 275-373\\
         8          & 0.6-0.8   & 385-500\\
         9          & 0.4-0.5   & 602-720 \\
         10         & 0.3-0.4   & 787-950 \\
         \hline \hline
    \end{tabular}
    \label{tab:alma}
\end{table}

    In Figure \ref{fig:spec_by_sci_theme}, we see a significant request for the atmospheric windows represented by ALMA bands 6, 7 and 8 for spectrometers ($211 - 500$ GHz) while Figure \ref{fig:cont_by_sci_theme} shows more of a concentration towards ALMA bands 3, 6 and 7 and an increase again for band 9 for continuum observations. This reflects the different types of science being achieved with these different types of cameras. 
    
    \begin{figure}[h]
         \centering
         \begin{subfigure}{0.45\textwidth}
             \centering
             \includegraphics[width=\textwidth]{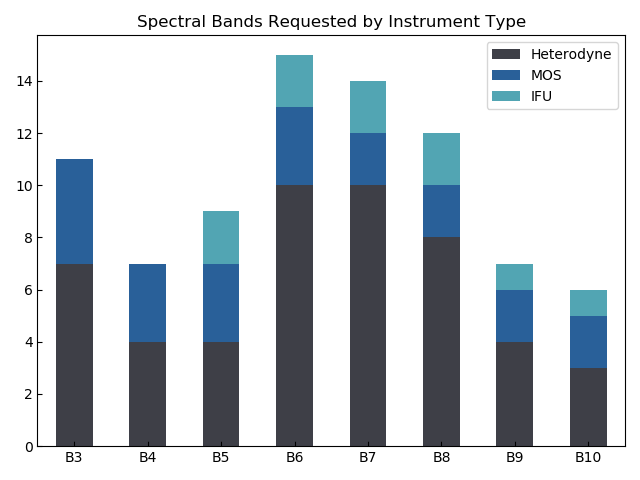}
             \caption{Types of Spectroscopic instruments requested by Science Theme}
             \label{fig:spec_inst_by_sci_theme}
         \end{subfigure}
         \hfill
         \begin{subfigure}{0.45\textwidth}
             \centering
             \includegraphics[width=\textwidth]{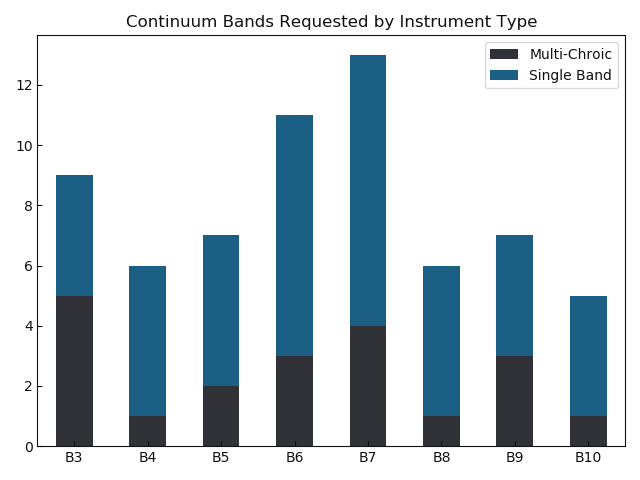}
             \caption{Types of Continuum observations requested by Science Theme}
             \label{fig:cont_inst_by_sci_theme}
         \end{subfigure}
            \caption{Spectroscopic and Continuum observations by broad instrument type as defined in the text. The labeling on the x-axis corresponds to the canonical ALMA bands\cite{cortes_paulo_2021_4612218} defined in each spectroscopic window.}
            \label{fig:Req_inst_by_theme}
    \end{figure}
    
    In Figure \ref{fig:Req_inst_by_theme}, we present the types of spectroscopic or continuum observations requested by science theme. 
    For spectroscopy, contributors were given examples of spectrometers such as a well sampled Heterodyne device, Integral Field Units (IFU) or Multi-Object Spectrographs (MOS). For continuum observations, with the understanding that there would be large format cameras, we asked whether the scientists required simultaneous multichroic observations, or whether single band observing would be sufficient. For most cases, we find a well sampled heterodyne or a single band continuum camera are sufficient for the general science cases, but note there are specific cases where these generalized instruments would be insufficient to meet the science requirements.    It is worth noting that the types of suggested instruments vary enormously in their technical readiness levels, and that respondents were therefore more likely to indicate instrumentation with which they are more familiar of the potential. As a result, this initial stage does not fully illustrate the demand for innovative instrumentation and further development of the science cases, along with instrumentation specialists, will be required to address this.
    
    In the following sections, we outline some of the key research questions in each of these four science themes, and the requirements imposed on the telescope by these science drivers.
    
    {\bf Together, the broad science coverage of the use cases, and the range of telescope characteristics that they require, shows that the community are looking for a multi-purpose facility with the potential to cover a wide range of science topics over decades to come.} 

        \subsection{The Sun and Solar System}
            
        The majority of the sub-mm emission from the Sun comes from the chromosphere, the interface layer between the photosphere and the corona, which plays a key role in energy and matter transport and is responsible for heating the outer layers of the solar atmosphere to temperatures over a million Kelvin. Studying this emission and understanding how its thermal and magnetic properties evolve over time will help us not only gain insight into the complex processes of energy transport but also solar activity and phenomena like solar flares. Understanding this in the Sun, where we can {\it resolve} the emission, will afford us a greater understanding of such processes in other stars, in turn allowing us to develop better methods of separating exoplanet signatures from their host star's ``background'' emission. Existing single dish submillimeter telescopes are unable to directly observe the Sun, or are very limited in resolution (e.g. the Solar Submillimeter Telescope\cite{2002RMxAC..14..149K}) but are key for the repeated observations necessary to study time-varying phenomena.
        
        As with the Sun, many planetary atmospheres in the Solar System are on large angular scales that become ``resolved out'' with interferometers like ALMA, making it impossible to study the temporal and spatial variations in these atmospheres on the Giant planets, Mars, and Venus with interferometers. Current single dish facilities are able to monitor the temporal variations, but the signal becomes beam diluted due the comparatively large beam sizes. Higher spatial resolution than currently available with single-dish facilities is also required to study icy moons; for instance, the angular resolution of modern facilities is of larger angular scales than the separation of Enceladus from Saturn (less than 10$''$ away), blending emission and making detailed study of Enceladus and Saturn's other moons extremely challenging (at 345 GHz, AtLAST would have a resolution of $\sim4''$).
        
        Comets, as extended sources on the sky, also suffer from their emission being filtered out by interferometers. By studying them with moderate resolution and great sensitivity, we should gain valuable insights on the prebiotic molecules which seed the Solar System and which could point towards the origins of life.
        
        Specific to Solar System science is the need for non-sidereal tracking: because Solar System bodies move with respect to background stars, the telescope will need to be able to track these non-sidereal objects. In addition to requiring high enough resolution to separate the atmospheres of planets and their icy moons, short cadence observations and rapid response to targets of opportunity (TOOs) will be required. The Sun rotates on a $\sim$ 1 month timescale and to capture a full solar rotation will require daily observations to properly quantify the Chromosphere. Planetary atmospheres vary on timescales of hours, and so repeated observations are required to capture that variability. If storms develop in a planetary atmosphere, the telescope needs to be able to respond quickly to that TOO.
        High resolutions of 1-2$''$ at the highest frequencies are required in order to spatially resolve the atmospheres of icy moons from those of their host planets.
        
        Both Solar and Solar System cases require broad spectral coverage to study the composition of the primordial (i.e. cometary) and modern day system. Broad spectral coverage will maximize the number of species detectable in a single observation, and also allows for line stacking for faint targets while minimizing calibration uncertainty.

        \subsection{The Milky Way}
        
        The science case for studying dust, chemistry, magnetism, dynamics and variability within our own galaxy primarily revolves around the formation of stars, and the gas and dust produced in evolved stars.
        
        Large-area studies of the Galactic Plane will constrain the initial mass function (IMF) and its relation to the core mass function (CMF) beyond the local (low-mass) censuses to date. They will allow us to constrain the dust opacity spectral index, the Galactic magnetic field, the transitions to and from molecular gas clouds, and the chemistry of the interstellar medium (ISM) in a global sense. With dedicated surveys of star forming regions, survey statistics will extend beyond the local star-forming regions in the Gould's Belt, to the more typical environments stretching from here to the Galactic Center. 
        
        Dedicated polarization studies of evolved stars will inform our understanding of dust and chemical evolution in these regions, and regular monitoring of all of these types of environments will further constrain variability in them, leading towards a better understanding of accretion and ejection events.
        
        To further constrain the evolution of our Galaxy, we can look to portions of the circumgalactic medium (CGM) which emit on such large scales that they are completely filtered out by interferometers. Studying, for instance, the Southern Fermi Bubble, extending $55^\circ$ from the galactic center, will allow us to quantify feedback from nuclear regions. Measuring the chemistry and kinematics of the Magellanic Stream, which stretches over $200^\circ$, is key to understanding the role of the tidal CGM. The faint, diffuse submillimeter emission from these regions is completely out of reach to current facilities, and at present the existence and strength of molecular gas tracers and cold dust in our own Galaxy is completely unknown. This reservoir of cold gas may play a crucial part in replenishing the ISM of the Milky Way, allowing for a smooth star formation history.
        
        Many of the Milky Way science cases require mapping of the Galactic Plane. To make these surveys feasible, the telescope needs to have a fast scanning mode. For example, for shallow, large area surveys, it is unlikely that the telescope will dwell for more than 10s at any point. With large format cameras, this necessitates fast scanning/slewing. A minimum continuum camera footprint of 10$'$ is required for such a camera in order to allow continuum observations to be combined with the larger scale images of facilities like {\it Planck}.
        
        To disentangle the dynamical and chemical complexity of the ISM requires high spectral resolution. Spectral resolutions better than 1 km/s allow us to quantify disk motions (often less than a few km/s) and spectrally disentangle emission lines with similar rest frequencies. This needs to tension with the large bandwidths required by many of the science cases (8-16 GHz) which will both capture the continuum emission in the line free regions, as well as minimize the number of passes required for full spectral coverage of each atmospheric window. These constraints, complemented by high spatial resolution, will lessen beam dilution in marginally resolved regions while simultaneously minimizing beam de-polarization.
        
        Simultaneous multi-chroic observations are needed for a number of cases to minimize the cross-band calibration uncertainties when building up Spectral Energy Distributions (SEDs) and to avoid washing out variability signatures by observing different bands at different times.
        
        To quantify variability in accretion and ejection mechanisms, we need repeat observations of statistical samples of objects, whether newly forming or evolved stars. The cadence of the repeats depends heavily on the science case, but can vary from weekly to yearly.

        \subsection{Nearby Galaxies}

        With the sensitivity and resolution achievable with a facility like AtLAST, evolved star studies comparable to those currently undertaken in the Milky Way can be undertaken in the Magellanic Clouds and beyond -- unlocking another dimension in the interplay between mass loss and chemical evolution because of their much lower metallicities.
        
        Similarly, CO and C{\sc I} surveys of nearby galaxies gives insights on the cold gas properties radiation fields, a robust determination of the CO to H$_2$ mass function, and the ability to study the interstellar medium (ISM) and CO-dark gas out to z$\sim$0.5. This then links directly to the diffuse CGM of these same galaxies, including molecular and atomic outflows and tidal tails. The molecular outflows of some nearby highly star-forming galaxies hosting obscured AGNs can embed up to $10^9$ -- $10^{10}$ M$_{\odot}$ of molecular gas, including high density gas. Their properties are not understood partially because it is hard to map their full extent beyond nuclear ISM scales with interferometers. The highly extended and diffuse emission from the intra-group and intra-cluster medium has been impossible to observe with existing facilities, and better understanding the physics of galaxy environments will help us determine the role these play in galaxy evolution.
                
        Many of the telescope requirements for the submitted Nearby Galaxy cases are very similar to those for our own Galaxy, focusing on spectral resolution and bandwidth.
        To quantify chemistry and dynamics in the ISM and CGM requires high spectral resolution, but with the caveat that binning up to 10 \kms{} could lower the overall spaxel count and therefore data rates.
        Stable spectral baselines are extremely important for studying the diffuse gas components of nearby galaxies\footnote{See the memo on a wobbler for AtLAST: \href{https://www.atlast.uio.no/documents/memo-series/memo-public/wobbler_for_atlast.pdf}{Potential wobbler solutions for AtLAST} by R.~Hills}.
        Larger bandwidths will drive down survey times for spectral scanning, but will also enable simultaneous observations of multiple species/isotopologues as has been shown with ALMA (i.e. simultaneous observations of $^{12}$CO in the upper sideband with $^{13}$CO and C$^{18}$O in the lower in ALMA band 6 ($\sim 211 - 275$ GHz).
        Bandwidths of at least 4 GHz at $\sim$ 900 GHz (ALMA Band 10) are crucial: in strong outflows, line widths can be broadened to up to 3000\kms, so wide instantaneous bandwidths are required to capture molecular emission lines in their entirety as well as sample adjacent line-free spectral regions for sufficient baseline measurements. Large bandwidths would also enable simultaneous observations of CO and C{\sc I} capturing both the CO-bright and CO-dark portions of Giant Molecular Clouds (GMCs) at the same time.
        
        To efficiently observe an entire galaxy cluster in a single pointing, and to minimize calibration uncertainty, requires a large FoV. As an example, a 1 sq. deg. FoV could be expected to simultaneously detect $\sim$ 40 galaxies with z$<$0.5.

        \subsection{Distant Universe and Cosmology}
        
        A large area spectroscopic and photometric survey at submillimeter wavelengths with the resolution, sensitivity and mapping speeds of AtLAST will overcome the confusion limit and allow us to probe ``normal'' star-forming galaxies, currently out of reach of single-dish observations, out to very high redshifts.

        AtLAST's wavelength coverage probes the peak of the dust spectrum of star-forming galaxies from z = 1 up to cosmic dawn and a plethora of spectral features therein.  With this information, we can probe strong cooling lines and map the reionization of the Universe, quantifying the properties of the very first galaxies. 
        With broad spectral coverage, we can determine redshifts and dust properties for a large sample of normal star-forming galaxies which  will enable the investigation of clustering of sources, further understanding of the role of environment in galaxy evolution, and quantifying the contributions of obscured star formation to the cosmic star formation rate density.
        Key to these goals is detecting samples of sources that are unbiased by obscuration out to high redshift, reaching much deeper than existing surveys with high flux confusion limits. 
        AtLAST's sensitivity to large scale, low surface brightness emission will allow studies of galaxy environment -- such as measuring the cold molecular gas in the CGM to investigate galaxy outflows, interactions, and inflows. 
        
        It will be possible to study the hot ($>10^6$~K) atmospheres of galaxy clusters and groups through their thermal SZ signatures, probing pressure, and through relativistic corrections to the thermal SZ effect, which constrains temperature. As the largest structures in the Universe, galaxy clusters and groups are sensitive to fundamental cosmological parameters and are excellent cosmological probes against which theory can be tested.
        
        The combination of arcsecond-scale angular resolution with multiple bands sampling the SZ effect spectrum opens up a multitude of possibilities for resolved SZ effect studies across a wide range of redshifts.  These possibilities include detecting shocks in merging galaxy clusters\cite{DiMascolo2019b}, probing exotic plasmas such as those found AGN bubbles\cite{Abdulla2019,Ehlert2019}, and exploring the variation in pressure profiles with redshift\cite{McDonald2017} and down to galaxy scales\cite{Amodeo2021}.  Above $\gtrsim 218$~GHz, AtLAST will enable constraints on temperature via the relativistic SZ effect, complementary to currently available X-ray information.
        
        Detailed measurements of submillimeter emission lines (i.e. their redshifts and line strengths) across large cosmological volumes will also allow us to map large scale structure using baryonic acoustic oscillations observable in matter clustering as a standard ruler to measure the expansion history of the Universe, and test models of dark energy and inflation and probe deviations from the $\Lambda$CDM model.
        
        In order to overcome cosmic variance and sample the large areas of sky required for measures of clustering on different cosmological scales, up to thousands of square degree sky coverage is required for a number of use cases, comparable to the SDSS, requiring multi-band photometry at sensitivities reaching the confusion limit.
        Fast mapping capabilities -- e.g. large FoV and high sensitivity -- make this possible, with efficiency improving if multichroic observations are possible simultaneously. Reaching a sensitivity of 0.1 mJy at ALMA band 6 ($\sim 211 - 275$ GHz) would result in detecting approximately 10$^{4}$ sources per square degree; a 1000 deg$^2$ survey area would thus reveal up to tens of millions of sources.
        All require the excellent resolution and sensitivity that a 50m dish could provide; however, constraints on spectral resolution are not as stringent as in other science themes:  only one submitted science case requires resolution $< 10$ km/s.
  
        Many of these science cases require 3+ bands of continuum observations over large sky areas. Simultaneous multi-band photometry would allow greater efficiency in sky surveys, but with the trade-off that this would mean fewer pixels at any given band. A large bandwidth is essential to capture spectral features over a broad redshift range, to maximize the cosmic volume probed by large area surveys, and will lower spectral scanning survey times.
        
        Two cases in particular make the case for sensitive, high-frequency ($>600$ GHz) observations.
        To resolve degeneracies between foreground dust components and SZ temperature signals, these high frequency bands are necessary. In modeling the dust properties of star-forming galaxies, high frequency observations are required to probe the peak of the dust spectrum at $z > 2$.
        
\section{Summary of key requirements and instrumentation}
\label{sec:requirements}

The use cases we received from the community present several pressing arguments that a 50m diameter telescope is required to push the boundaries of understanding as it breaks through a number of barriers imposed by current observational facilities. This diameter is fairly consistent with the longest baselines of the ALMA compact Array (ACA, 45m baselines), however with greater sensitivity and throughput, AtLAST opens up the  ability to quickly scan large portions of the sky, while maintaining sensitivity to large scale structure.

Table \ref{tab:requirements_overview} presents an overview of the range of telescope properties requested in the submitted cases.  In all science themes, polarimetric observations are required, and for the more nearby objects, repeat observations with specific cadences are required to capture time variability on timescales of days to months. Additionally, as we have framed the use cases as science-motivated rather than by observational strategy, we do not include the multitude of possible Very Long Baseline Interferometry (VLBI) science cases yet; the potential of AtLAST to participate in VLBI activities is still to be explored.

In terms of spatial sampling Table \ref{tab:requirements_overview} presents two options: maps that when properly observed are fully sampled, and those that pick off specific parts of the FoV to sample. These distinctions came primarily from spectroscopic observations where the number of individual pixels is much more limited than from continuum cameras.  

In the case of fully sampled mapping, this does not necessarily imply fully sampled cameras. Sparsely sampled detectors can generate fully sampled maps when the right mapping methods are applied. For the MOS observations, the concept includes placing a number of fibres (or an IFU) at each position so that a number of small `postage stamp' maps are created within a single telescope FoV.  

\begin{table}[hbt!]

    \caption{Generalized telescope requirements pulled from the submitted use cases. The Atmospheric windows are listed with respect to the ALMA observing bands, for which a translation to wavelength/frequency can be found in the ALMA technical handbook\cite{cortes_paulo_2021_4612218} and for reference in Table~\ref{tab:alma}. In the final row showing continuum resolution required, ``1/Band'' means 1 channel per band.}
    \centering
    \begin{tabular}{lr|cccc}
    \hline \hline
         & & Solar System & Milky Way & Nearby Galaxies & Distant Universe  \\
    \hline
    Diameter & (m) & 50 & 50 & 50 & 50\\
     \multicolumn{2}{l|}{Instantaneous Field of View  (deg$^2$)} & - & $>$ 0.03 & 1 & - \\
    Largest Angular Scale & (arcmin) & $\sim$ 30 & 10 & - & 60 \\
    
    & \\
    \multicolumn{2}{l|}{Atmospheric Windows} &  B3-B10 & B3-B10 & B1-B10 & B1-B10 \\
    Spatial Coverage & (deg$^2$) & 0.5 &  540 & 1000+ & 1000+ \\
    Sampling & & Full & Full/MOS & Full & Full \\
    & \\
    Repeats && \checkmark{} & \checkmark{} & \checkmark{} & \checkmark{} \\
    Time Variability & & & \checkmark{}  & \checkmark{} \\
    \multicolumn{2}{l|}{Targets of Opportunity} & \checkmark{}  & \checkmark{} \\
    Fixed Schedule & & \checkmark{} & \checkmark{} & & \checkmark{} \\
    Time Critical &&  \checkmark{} \\
    &\\
    Polarization & & \checkmark{} & \checkmark{} & \checkmark{} & \checkmark{} \\
    &\\
    Spectral Bandwidth & (GHz) & 4-16 & 8-16 & 16+ & 16+ \\
    Spectral Resolution & (km/s) & 0.1 & 0.05 & 0.1 & 1\\
    &\\
    Continuum Bandwidth & (GHz) & 16+ & 16+ & 16+ & 16+\\
    Continuum Resolution & (km/s) & 1/Band & 1/Band  & 1/Band  & 1/Band \\
    
    \hline\hline
    \end{tabular}
    \label{tab:requirements_overview}
\end{table}

    \subsection{Expected Instrumentation suite}
    Within the submitted use cases, themes emerged with respect to the types of instrumentation required by the science. 
    The generalized instrumentation suite from this exercise comprises: a large format continuum camera\cite{2018SPIE10708E..0JB}, a highly multiplexed heterodyne~\cite{2019arXiv190703479G}, and a multi-object spectrometer, comparable to current state-of-the-art technology but thinking ahead to next-generation instruments.  Overall, cases were made for an ultra-wide bandwidth IFU~\cite{2016JLTP..184..114B}, however the support for such an instrument was much lower than for the other spectrometers.  
    As mentioned earlier, the suggested cases lean towards instrumentation similar to existing facilities, and further work needs to be done to investigate the potential of novel technologies.
    For large format continuum cameras, there was little variation between requested capabilities, especially when considered against the spectrometer requests; the most notable variations being that the requests for a multi-object spectrometer were more numerous than for an IFU.
    
        \subsection{Continuum Camera}
        
        As shown in Figure \ref{fig:cont_by_sci_theme}, there is the desire to cover a broad range of (sub-)mm windows with a continuum camera. The bulk of the requests focus on the 1mm (300 GHz) regime covering ALMA bands 6 and 7. With the proposed site options at Chajnantor, AtLAST would be very well placed to observe at these frequencies. However it is clear from the distribution of requests that scientists want to probe as much of the dust continuum SED as they can, across as many redshifts as possible, and it is this access to high frequencies that necessitates the proposed altitude and location of AtLAST: elsewhere, these bands would be inaccessible.
        
        The general consensus was for at least 16 GHz of instantaneous bandwidth, with some science cases requiring that to be broken down into channels of order 1 MHz to allow for the removal of spectral line contamination.
        
        The required angular resolution for the individual science cases is consistent with that of a 50m primary mirror, with only a few cases requiring the 1-2$''$ precision of the highest frequency observations. The largest angular scales required of the continuum camera were of order 60$'$, with many of the surveys requesting upwards of 1000 sq. deg for their overall footprints.
        
        Studies of Solar System objects and protostellar variability place constraints on being able to simultaneously observe in different wavebands which not only minimizes uncertainty between observing bands, but allows for the SEDs of the objects to be timestamped together: variability on day-to-day timescales can then be attributed to astrophysics, rather than calibration uncertainty.
        
        \subsection{Spectrometers}
        
        For spectroscopic observations, there was a much clearer drive towards the 1mm (300 GHz) and shorter wavelength bands than in the continuum case, especially in the nearby Universe where the CO (J=3-2, J=2-1) and [CI] ($^3P_1 - ^3P_0$) transitions fall.  For the distant Universe, there is a much broader request for virtually all bands, as emission is redshifted to lower frequencies -- enabling detection of rest-frame IR and optical lines.
        
        Many of the chemical surveys require full coverage of each (sub-)mm atmospheric window, with broad bandwidths (e.g. 8-16 GHz) to minimize the number of repeat observations required to fully sample the window. With requested spectral resolutions down to 0.1 km s$^{-1}$, this poses a technological challenge for both data collection and processing.
        
        The largest angular scales requested are of order 0.03 degree (for observations of, for example, the Sun), with resolutions enabled by a 50m diameter dish.
        
        There were few requests for simultaneous multi-band observations, rather there is a range of opinions on how different users would want to use the large focal plane. First generation heterodynes will not be able to fill AtLAST's 2 degree FoV, and so below we discuss the two observing modes most commonly requested for spectroscopic observations.
        
        \subsubsection{Highly Multiplexed Spectrometer}
        
        To maximize the efficiency of the spectroscopic surveys, a large format heterodyne is required. Being able to quickly and efficiently obtain fully sampled maps of a region (whether it be the Galactic Plane or a cosmological survey field) will provide excellent statistics on the (sub-)mm sky.  This means of order 1000 pixels or more are needed for the spectrometer.

        \subsubsection{Multi-Object Spectrometer}
        
        Not all science cases require a fully sampled map. In these cases, small clusters of spectrometers (or small IFUs) that are positionable throughout AtLAST's large FoV would allow studies of smaller objects, like individual galaxies in nearby clusters or protostellar cores in a star-forming region.
    
    \subsection{Other instrumentation requests}
    
    The generalized instrumentation above includes the parameters given for solar observing, however we note that instruments required to observe the Sun will require special coatings and considerations that were beyond the scope of this use case exercise. These parameters will be drawn out in more detail in the following steps of the science case development.
    
    There were a few requests for ultra-wideband heterodyne receivers, primarily the blind surveys of the distant Universe.  Here, the science cases requested the full atmospheric window across the full sub-mm range for line-intensity mapping studies at z $>$ 2.8.

\section{Next steps}
\label{sec:future}
The results presented here are only the first steps in defining and developing the science cases. For the remainder of the study, we will focus on building robust, quantified justification for strong science drivers answering key questions in the next decades of submillimeter astrophysics that only a high-throughput, sensitive single dish telescope like AtLAST can achieve.

To this end there are a number of threads in work going forward. Already in development is a sensitivity calculator to aid in estimating exposure times for individual suggested observations for science cases. This software will allow us to quantify time estimates for point-source sensitivities, putting constraints on the feasibility of science cases. There is also work underway dedicated to modeling the sub-mm line emission of the ISM and CGM, using high resolution zoom-in simulations (such as PONOS\cite{2017MNRAS.467.4080F}), which will be followed up with a similar study of SZ effect observables in future.

The initial science cases are being developed further into thematic case studies by collaborative groups including both internal members of the AtLAST consortium and the wider astronomy community. The aim is that by the end of the design period, we have quantified and consistent case studies on each science theme that outline the key science questions that AtLAST will be able to address. The results of these case studies will be used to populate a matrix of requirements for the telescope. This and other memos will become available on the AtLAST website over the course of the design study\footnote{\href{https://www.atlast.uio.no/}{https://www.atlast.uio.no/}}. The final deliverable of the science work package (due 2024) is the publication of a science overview report, which will present a compelling argument for the need for a telescope, and illustrate that across the astrophysics community there is a clear demand for a multi-instrument, high throughput, community facility to address a range of fundamental questions about the submillimeter Universe.  For information on how to contribute your own science case, please visit the AtLAST website.

\acknowledgments 
We are extremely grateful for the guidance from, and the tireless devotion of, Richard Hills throughout the AtLAST optical design.
This project has received funding from the European Union's Horizon 2020 research and innovation programme under grant agreement No 951815.

\bibliography{report} 
\bibliographystyle{spiebib} 

\end{document}